\documentclass[aps,twocolumn,showpacs]{revtex4}

\usepackage{amsmath}
\usepackage{bm}
\usepackage{graphicx}

\newcommand{\ket}[1]{\, | #1 \rangle}
\newcommand{\braket}[2]{\langle #1 | #2 \rangle}
\newcommand{\expv}[1]{\langle #1 \rangle}
\newcommand{\bk}{b^{\dagger}}
\newcommand{\ck}{c^{\dagger}}
\newcommand{\hn}{\hat{n}}
\newcommand{\hmd}{\hat{m}}
\newcommand{\om}{\omega}
\newcommand{\Om}{\Omega}
\newcommand{\be}{\begin{equation}}
\newcommand{\ee}{\end{equation}}
\newcommand{\bea}{\begin{eqnarray}}
\newcommand{\eea}{\end{eqnarray}}
\newcommand{\besa}{\begin{subequations}\begin{eqnarray}}
\newcommand{\eesa}{\end{eqnarray}\end{subequations}}
\newcommand{\hlf}{\mbox{$\frac{1}{2}$}}  

\begin{document}

\title{Quantum dynamics of one and two bosonic atoms 
in a combined tight-binding periodic and weak parabolic potential}

\author{Manuel Valiente}
\affiliation{Institute of Electronic Structure \& Laser, FORTH, 71110
Heraklion, Crete, Greece}
\author{David Petrosyan}
\affiliation{Institute of Electronic Structure \& Laser, FORTH, 71110
Heraklion, Crete, Greece}

\date{\today}

\begin{abstract} 
Strongly interacting bosonic particles in a tight-binding periodic potential
superimposed by a weak parabolic trap is a paradigm for many cold atom 
experiments. Here, after revisiting the single particle problem,
we study interaction-bound dimers of bosonic atoms in the combined lattice
and parabolic potential. We consider both repulsively- and attractively-bound
dimers and find pronounced differences in their behaviour. We identify 
conditions under which attractive and repulsive dimers exhibit analogous 
dynamics. Our studies reveal that coherent transport and periodic
oscillations of appropriately prepared one- and two-atom wavepackets can 
be achieved, which may facilitate information transfer in optical 
lattice based quantum computation schemes.
\end{abstract}


\maketitle

\section{Introduction}

Quantum transport in periodic structures is one of the central topics 
of condensed matter physics \cite{SolStPh}. Recent interest towards 
spatially-periodic systems has been largely motivated by the remarkable
progress in cooling and trapping bosonic and fermionic atoms in optical 
lattices \cite{OptLatRev}. The relevant parameters of these systems 
can be controlled with very high precision and can be tuned to 
implement some of the fundamental models of condensed matter physics.
In a tight-binding regime, the Hubbard model accurately describes
static and dynamic properties of these systems \cite{OptLatRev,JakZol}.
Importantly, in real experiments with cold atoms, the lattice is finite 
and often it is superimposed by a weak harmonic trap. This breaks the 
translational invariance of the lattice and thereby strongly modifies 
the properties of the system even in the limit of non-interacting 
particles \cite{HooQui,RPCW,rigol}. In particular, the low energy 
states of the single-particle spectrum behave like harmonic oscillator 
states while the higher energy states are localized at the sides 
of the parabolic trap, which can lead to inhibition of quantum 
transport and dipole oscillations in a degenerate atomic gas 
\cite{rigol,njptr,inguscioLoc,fertig}. 

Atom-atom interactions profoundly enrich the Hubbard model, 
as attested by, e.g., theoretical prediction followed by spectacular 
experimental demonstration of the transition from the superfluid 
to the Mott insulator phase in an ensemble of cold bosonic atoms 
in optical lattice \cite{optlattMI}.
Strongly interacting bosons in periodic potentials can form 
tightly bound ``dimers'' \cite{PSAF} observed in a recent experiment 
\cite{KWEtALPZ} with repulsively interacting atoms in a lattice. 
Here we first discuss static and dynamic properties of a single
atom in a combined periodic and weak parabolic potential and show 
that coherent transport and perfectly periodic oscillations of 
appropriately prepared wavepackets can be achieved. We then study
the properties of strongly interacting atom pairs examining both 
regimes of attractive as well as repulsive interactions. We show 
that, quite generally, the interaction-bound dimers behave as 
single particles with appropriately rescaled parameters of the 
system. We identify, however, important differences between  
attractively-bound and repulsively-bound dimers and find, 
rather surprisingly, that, in a weak trap, the repulsive dimer 
is bound stronger than the attractive one.

\section{The model}

We consider cold bosonic atoms in a combined tight-binding periodic 
and weak parabolic potential. In 1D, the system is described by the 
Bose-Hubbard Hamiltonian
\be
H = \sum_{j} \left[ \Om j^2 \hn_j + \frac{U}{2}  \hn_j (\hn_j-1)
-J (\bk_j b_{j+1} + \bk_{j+1} b_j ) \right] \, , \label{BHHam}
\ee
where $\bk_{j}$ ($b_{j}$) is the creation (annihilation) operator 
and $\hn_j = \bk_{j} b_{j}$ the number operator for site $j$, 
$J$ is the tunnel coupling between adjacent sites, $U$ is the 
on-site interaction, and $\Om > 0$ quantifies the strength of 
the superimposed parabolic potential due to which site 
$j = \pm 1, \pm 2, \ldots$ acquires energy offset $\Om j^2$ with 
respect to site $j = 0$ corresponding to the minimum of the potential.
A natural basis for Hamiltonian (\ref{BHHam}) is that of the 
eigenstates $\ket{n_j} \equiv \frac{1}{\sqrt{n !}} (\bk_j)^{n} \ket{0}$
of operator $\hn_j$ whose eigenvalues $n = 0,1,2,\ldots$ denote the 
number of particles at site $j$, and $\ket{0} \equiv \ket{\{ 0_j \}}$ 
is the vacuum state.

%\subsection{Single particle spectrum}
\paragraph*{Single particle spectrum.}

\begin{figure}[tb]
\includegraphics[width=0.38\textwidth]{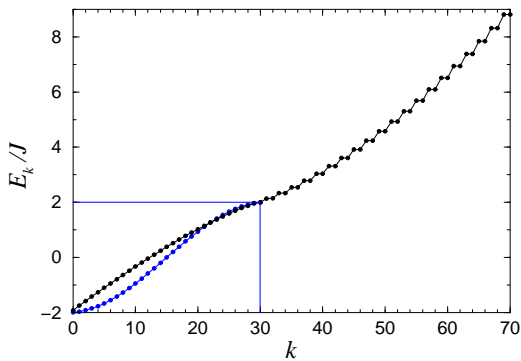}\\
\caption{Single particle energy eigenvalues $E_k$ in a combined 
periodic and parabolic potential obtained by numerical diagonalization 
of Hamiltonian (\ref{BHHam}) with $J/\Om = 140$. For comparison, the blue 
dots represent eigenvalues $\bar{E}_k$ within the Bloch (mini)band for a 
flat lattice ($\Om = 0$) of length $\bar{N} = N = 31$.}
\label{fig:EnHub}
\end{figure}

We first discuss the single-particle case with the on-site interaction 
$U$ playing no role. Recall that in the absence of parabolic potential, 
$\Om =0$, the eigenstates of the Hubbard Hamiltonian (\ref{BHHam}) form 
a Bloch band of width $4J$ centered around zero. More quantitatively, 
given a finite flat lattice of $\bar{N}$ sites, the Bloch eigenenergies
and corresponding eigenstates are given by 
\besa
\bar{E}_k &=& - 2J \cos \left[ \frac{\pi (k + 1)}{\bar{N}+1} \right] \, , 
\label{EfinBB} \\
\ket{\bar{\chi}_k} &=& \mathcal{N}  
\sum_{l=1}^{\bar{N}} \sin \left[ \frac{l \pi (k + 1)}{\bar{N}+1}\right]  
\ket{1_l} \, ,   \label{psifinBB} 
\eesa
with $0 \leq k < \bar{N}$ and $\mathcal{N}$ a normalization constant. 
In the limit of $\bar{N} \to \infty$, Eq.~(\ref{EfinBB}) yields the 
well-known dispersion relation \cite{SolStPh} 
$\bar{E}_q =  - 2 J \cos(q)$ with $0 \leq q \leq \pi$ and lattice 
constant $d=1$. Remarkably, however, even a very weak parabolic
potential $\Om \ll J$ drastically modifies the spectrum of 
Hamiltonian (\ref{BHHam}) \cite{RPCW,HooQui}, as shown in 
Fig.~\ref{fig:EnHub}. Note that the spectrum, bound from below by $-2 J$, 
is composed of discrete energy levels $E_k$. Two distinct groups of levels
can be identified: (i) the low-energy levels $E_k \leq 2 J$ forming a 
modified Bloch band, and (ii) the high-energy ones $E_k > 2 J$. 

(i) The parabolic potential effectively restricts the number of sites 
accessible to a particle with energy within the Bloch band 
$-2 J \leq E_k \leq 2 J$. Roughly, only sites $j = 0, \pm 1, \ldots$, 
for which $\Om j^2 < 2 J$, can participate in the formation of the 
low-energy part of the spectrum \cite{RPCW}. More precisely, using 
second order perturbative corrections, we find that the modified Bloch 
band is restricted to sites $j$ satisfying
\be 
|j| \leq j_{\textrm{max}} \equiv \sqrt{\left( 1 + \frac{1}{\sqrt{2}} \right) 
\frac{J}{\Om} } \simeq 1.3 \sqrt{\frac{J}{\Om} }  \, .
\ee 
The low-energy part of the spectrum therefore contains 
$N = 2 \lfloor j_{\textrm{max}} \rfloor + 1$ energy levels 
$E_0,E_1, \ldots E_{N-1}$, which is illustrated in Fig.~\ref{fig:EnHub}, 
and we have verified this conclusion for a wide range of values of $J/\Om$. 
Note that the weak parabolic potential modifies the Bloch band in such a 
way that its lowest energy part is approximately linear in $k$, similar 
to the spectrum of a harmonic oscillator. Indeed, using the properties 
of the Mathieu functions \cite{{abramowitz}}, it can be shown \cite{RPCW} 
that in the limit of $J/\Om \gg 1$, the low-energy eigenvalues $E_k$ and 
eigenstates $\ket{\chi_k}$ of Hamiltonian (\ref{BHHam}) are well 
approximated by 
\besa
E_k &\approx& -2J + 2 \sqrt{J \Om} \, ( k + \hlf ) \, , \label{eigEnk}  \\
\ket{\chi_k} &\approx& \mathcal{N} 
\sum_j (2^k k!)^{-1/2} e^{- \zeta_j^2/2 } H_k(\zeta_j) \ket{1_j} \, ,  
\label{eigStk}
\eesa
where $\mathcal{N}$ is a normalization constant, 
$\zeta_j = j \sqrt[4]{\Om /J}$ is the discretized coordinate, 
and $H_k(\zeta)$ is the $k$-th Hermite polynomial. 
Thus, the lowest energy eigenvalues $E_k$ and eigenstates $\ket{\chi_k}$
correspond to those of an energy-shifted harmonic oscillator with an 
effective frequency $\hbar \om =  2 \sqrt{J \Om}$ and an effective
mass given by the usual expression $\mu = \hbar^2/(2J d^2)$ \cite{SolStPh} 
which is valid near the bottom of the lowest Bloch band of a periodic 
potential. In particular, the ground state with $E_0 = -2J + \sqrt{J \Om}$
is given by
\be
\ket{\chi_0} = \sqrt[8]{\frac{\Om}{\pi^2 J}}
\sum_j e^{- \zeta_j^2/2 } \ket{1_j} \, . \label{grSt} 
\ee
The spectrum at the top of the modified Bloch band approaches
that of a uniform ($\Om =0$) finite ($\bar{N} = N$) lattice given by 
Eq.~(\ref{EfinBB}). 

(ii) On the other hand, the high-energy eigenvalues $E_k > 2 J$ with 
$k \geq N$ are two-fold degenerate, as seen in Fig.~\ref{fig:EnHub}. 
The pairs of degenerate states with indices $k = 2|j| +1$ and 
$k' = 2|j| + 2$ are localised around sites $j = \pm |j|$ 
($|j| > j_{\textrm{max}}$) equidistant from the center of the 
parabolic potential, the corresponding energies being given by 
$E_{k,k'} \approx \Om j^2$ \cite{RPCW,HooQui}. For such states, 
the localization occurs because, for large enough $|j|$, the 
transitions $\ket{1_j} \to \ket{1_{j\pm 1}}$ effected by the last 
term of Hamiltonian~(\ref{BHHam}) become non-resonant and the particle 
tunneling between neighboring lattice sites is suppressed. These 
high-energy states were shown to be responsible for damping of 
dipole oscillations and quantum transport of degenerate atomic 
gases in combined harmonic and optical lattice potentials 
\cite{inguscioLoc,rigol,fertig,RPCW,njptr}. Such localised states 
can be selectively addressed by radio-frequency fields \cite{inguscioAdr} 
and may be employed for efficient initialization of a qubit register 
with fermionic atoms \cite{smerzi}.

%\subsection{Coherent dynamics of a single particle wavepacket}
\paragraph*{Coherent dynamics of a single particle wavepacket.}

\begin{figure}[t]
\includegraphics[width=0.42\textwidth]{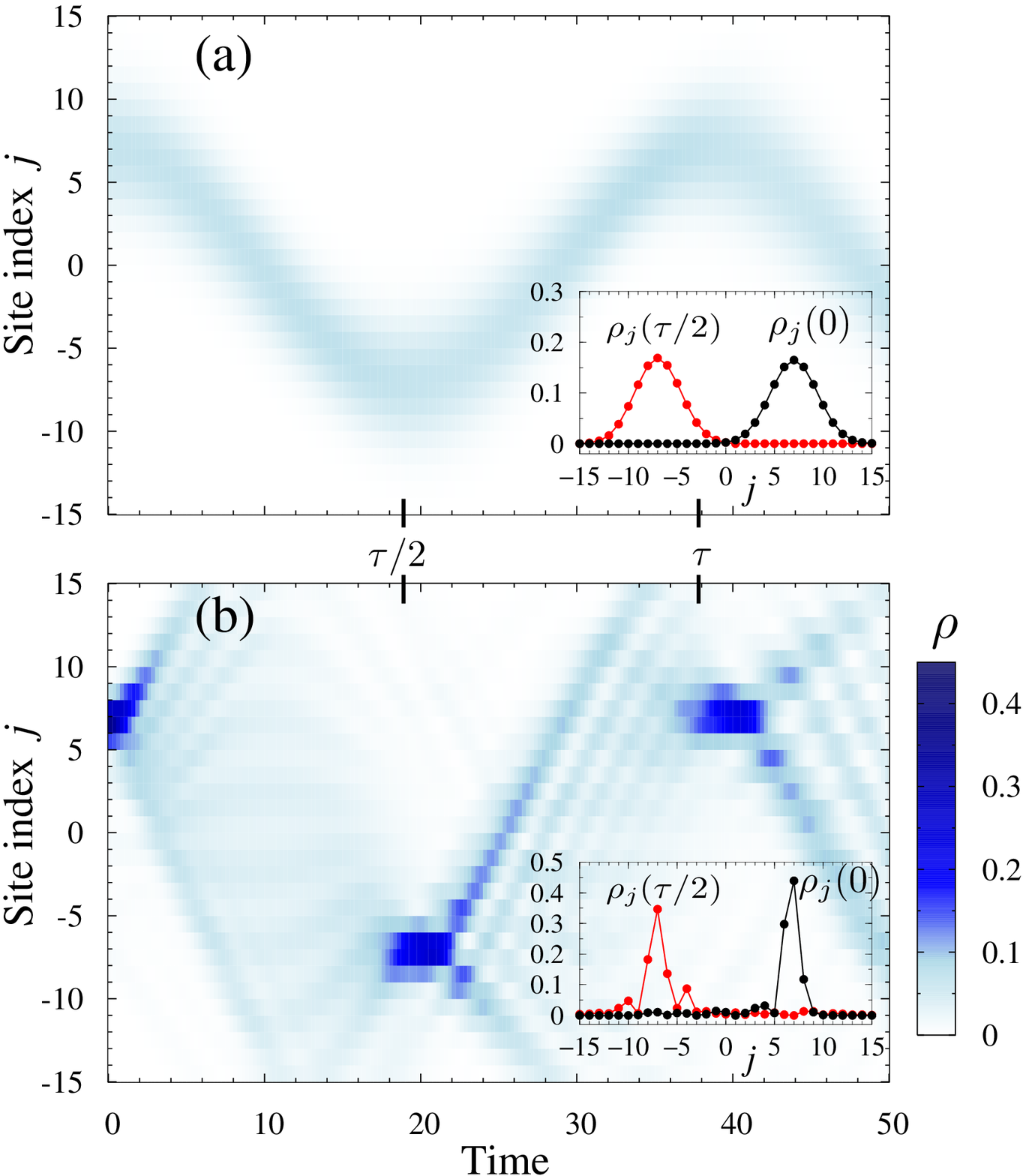}
\caption{Time evolution of density $\rho_j \equiv \expv{\hn_j}$ for 
a single particle wavepacket $\ket{\psi}$ in a combined periodic 
and parabolic potential with $J/\Om = 140$. 
(a) Initial state $\ket{\psi(0)}$ corresponds to the ground state 
$\ket{\chi_0}$ (discrete Gaussian) shifted by 7 sites from the trap center. 
(b) Initial state $\ket{\psi^{(j')}(0)}$ is a localized around $j' = 7$ 
wavepacket constructed from the $k=0,1,\ldots 20$ eigenstates $\ket{\chi_k}$.
Insets in (a) and (b) show the density distribution $\rho_j$ at $t = 0$ 
and $t \simeq \tau/2$. Time is measured in units of $\hbar J^{-1}$.}
\label{fig:spdyn}
\end{figure}

From the above analysis, it is clear that if we restrict ourselves to 
the harmonic oscillator--like states belonging to the lowest part of 
the energy spectrum, we can expect a quasi-periodic dynamics in the system.
Non-dispersive transport of a single particle wavepacket from one side 
of the shallow parabolic potential to the other can then be achieved. 
In Fig.~\ref{fig:spdyn}(a) we show the dynamics of a single particle 
wavepacket $\ket{\psi}$, represented by the ground state of the system 
$\ket{\chi_0}$, Eq.~(\ref{grSt}), initially shifted by $7$ sites from 
the trap center. Numerical solution of the Schr\"odinger equation using 
Hamiltonian (\ref{BHHam}) reveals almost perfect periodic oscillations 
of the discrete Gaussian wavepacket between the two sides of the parabolic 
potential with period $\tau \simeq 2 \pi /\om = (\pi \hbar/ J)  \sqrt{J/ \Om}$. 

From the set of harmonic oscillator--like states $\ket{\chi_k}$
of Eq. (\ref{eigStk}), we can construct a well-localized wavepacket
$\ket{\psi^{(j')}}$ centered at a prescribed site $j'$ 
($|j'| < j_{\textrm{max}}$). If we write the initial state as
\be
\ket{\psi(0)} = \sum_k A_k \ket{\chi_k} \, , \label{spLocSt}
\ee 
the probability amplitude $a_j$ for a particle to be at site $j$ is given by 
\be
a_j = \braket{1_j}{\psi(0)} 
\propto \sum_k A_k \, (2^k k!)^{-1/2} \, e^{- \zeta_j^2/2 } \, H_k(\zeta_j) \, . 
\ee
To obtain a localized around site $j'$ state $\ket{\psi^{(j')}}$, we 
maximize $|a_{j'}|^2$, which determines the set of coefficients $\{ A_k \}$
in Eq.~(\ref{spLocSt}). In Fig.~\ref{fig:spdyn}(b) we show the time 
evolution of such a localized state, which exhibits periodic collapses 
and partial revivals at sites $-j'$ and $j'$ with time steps $\tau/2$.  
The revivals are not complete since, as noticed above, the energy
spectrum $E_k$ for small $k$ is only approximately linear in $k$.
Nevertheless, our results suggest that coherent non-dispersive transport 
of carefully engineered atomic wavepackets can be achieved in  
optical lattices in the presence of a shallow parabolic potential. 

\section{Two particle dynamics}

\begin{figure}[t]
\includegraphics[width=0.42\textwidth]{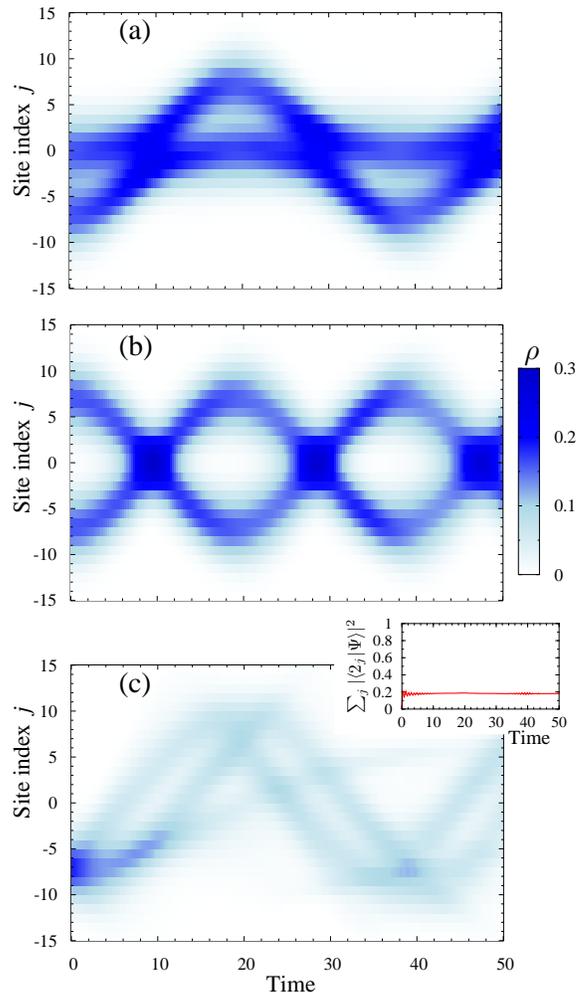}
\caption{Time evolution of density $\rho_j \equiv \expv{\hn_j}$ 
for two particles in a combined periodic and parabolic potential with 
$J/\Om = 140$ and $U = -10 J$. 
(a) Initial state $\ket{\Psi(0)}$ corresponds to one particle 
in the ground state $\ket{\chi_0}$ and the other particle in state 
$\ket{\chi_0}$ shifted from the trap center by 7 sites. 
(b)~Initially both particles in state $\ket{\chi_0}$ are shifted 
from the trap center by 7 sites in opposite directions. 
(c)~Initial state corresponds to both particles in state 
$\ket{\chi_0}$ shifted from the trap center by 7 sites in the same direction. 
Inset in (c) shows the projection $\sum_j |\braket{2_j}{\Psi}|^2$.}
\label{fig:tnipdyn}
\end{figure}

We now consider two bosonic particles in the combined periodic 
and weak parabolic potential. Clearly, in the simplest case of 
feeble interaction $|U| \ll J$, we have two independent 
particles for which the results of the previous section apply. 
But even for strong on-site interaction $U$, some aspects of 
the combined dynamics of two low-energy particles can be 
inferred from the independent particle picture modified 
by short-range collisions. This applies when the initial 
state $\ket{\Psi(0)} = \ket{\psi} \otimes \ket{\psi^{\prime}}$ 
is composed of two non-overlapping single-particle wavepackets, 
$|\braket{\psi}{\psi^{\prime}}|^2 \ll 1$, which upon collision 
with each other are reflected by the potential barrier 
$|U| \gtrsim J$. Examples of such a situation with large 
on-site attractive interaction energy $U = -10 J$ are shown 
in Figs.~\ref{fig:tnipdyn}(a) and \ref{fig:tnipdyn}(b). 
Analogous dynamics is observed for the repulsive  
interaction $U = 10 J$.

More intriguing is the case of initial state 
$\ket{\Psi(0)} = \ket{\psi} \otimes \ket{\psi}$ 
consisting of two overlapping single-particle wavepackets 
shown in Fig.~\ref{fig:tnipdyn}(c). This state has a significant 
population of the two-particle states $\ket{2_j}$ given by 
$\sum_j |\braket{2_j}{\Psi}|^2 \simeq \sum_j |a_j|^{4}$, 
where $a_j$ are the single-particle probability amplitudes. 
Clearly, the population of two-particle states is largest 
in the central part of the initial density distribution. 
As seen in Fig.~\ref{fig:tnipdyn}(c), this part exhibits 
slow dynamics, characterized by the effective tunnelling 
constant $J^{(2)} = - 2J^2/U$ (see below), and separates from 
the wings of the initial density profile. The wings, formed by 
the single-particle states $\ket{1_j}$, oscillate between the 
two sides of parabolic potential with the usual period $\tau$. 

%\subsection{Interaction-bound dimers}
\paragraph*{Interaction-bound dimers.}

At this point, let us recall \cite{KWEtALPZ,PSAF} that two bosonic
particles occupying the same site $j$ can form an effective ``dimer'' 
bound by the on-site interaction $U$. Thus, when $|U| \gg J$, the first-order 
transitions $\ket{2_j} \to \ket{1_j} \ket{1_{j\pm 1}}$ effected by the 
last term of Hamiltonian (\ref{BHHam}) are non-resonant and the 
particles can not separate. However, the second-order in $J$ transitions
$\ket{2_j} \to \ket{2_{j\pm 1}}$ via virtual intermediate states 
$\ket{1_j} \ket{1_{j\pm 1}}$ are resonant. Consequently, the dimer can 
tunnel as a whole with the effective rate $J^{(2)} = - 2J^2/U \ll J$ 
\cite{PSAF}. This explains the dynamics seen in Fig.~\ref{fig:tnipdyn}(c)
where the initial density distribution splits into slow and fast 
propagating components, the former composed of the dimer states 
$\ket{2_j}$ while the latter containing the monomer states $\ket{1_j}$. 

If the initial state is prepared in such a way that only two-particle
(dimer) states are populated, as implemented in, e.g., \cite{KWEtALPZ}, 
for $|U| \gg J $ the system can, to a good approximation, be described 
by an effective dimer Hamiltonian derived in the second order in $J/U$
\cite{PSAF}. In terms of the dimer creation 
$\ck_j= (\bk_j)^2 [1/\sqrt{2(\hn_j+1)}]$ and 
annihilation $c_j= [1/\sqrt{2(\hn_j+1)}] (b_j)^2$ operators, and 
number operator $\hmd_j = \ck_j c_j = \hn_j/2$, the effective Hamiltonian 
for a single dimer reads 
\footnote{For the sake of convenience, here the sign of the dimer tunnel 
coupling $J^{(2)} = - 2 J^2/U$ is taken opposite to that in \cite{PSAF}, 
while the nearest neighbour interaction does not play a role since 
we consider only a single dimer in the system.}
\be
H_{\mathrm{eff}} = \sum_{j} \big[\Om^{(2)} j^2 \hmd_j + (U - J^{(2)} ) \hmd_j 
-J^{(2)} (\ck_j c_{j+1} + \ck_{j+1} c_j ) \big] \, , \label{DimHam}
\ee
where $\Om^{(2)} = 2\Om$ is the strength of a parabolic potential 
seen by the dimer, while $(U - J^{(2)})$ represents the ``internal'' 
energy of a dimer. 

Before proceeding, let us note that, differently from the flat lattice 
situation considered in \cite{PSAF}, here the effective Hamiltonian 
$H_{\mathrm{eff}}$ is not applicable in the vicinity of sites 
$|j| \simeq |U|/(2\Om)$ where near-resonant dissociation of a dimer 
can occur via transitions $\ket{2_j} \to \ket{1_j}\ket{1_{j\pm 1}}$.
But since we are interested in the dynamics of low-energy dimers
with $|U| \gg J \gg \Om$, such high-$j$ states cannot be reached.  

\begin{figure}[t]
\includegraphics[width=0.42\textwidth]{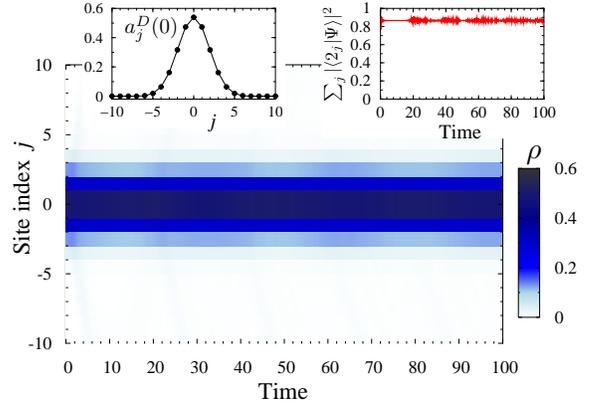}
\caption{Time evolution of atom density $\rho_j \equiv \expv{\hn_j}$ 
for an attractively-bound dimer in a combined periodic and parabolic 
potential with $J/\Om = 140$ and $U = -10 J$.  
Initial state $\ket{\Psi(0)}$ corresponds to the ground state
$\ket{\chi_0^{D}}$ of the effective Hamiltonian~(\ref{DimHam}) 
with the dimer amplitudes $a_j^D(0)$ shown in the left inset. 
Right inset is the projection $\sum_j |\braket{2_j}{\Psi}|^2$.}
\label{fig:adimdyn}
\end{figure}

Consider first the case of strong attractive interaction $U<0$ leading
to a positive tunnelling constant $J^{(2)} > 0$. Then the effective 
Hamiltonian (\ref{DimHam}) has the same form as the Hubbard Hamiltonian
(\ref{BHHam}) for a single particle in a combined periodic and parabolic
potential. We can therefore immediately write the lowest energy 
eigenvalues and eigenstates for an effective dimer as
\besa
E_k^D &\approx& -2J^{(2)} + 2 \sqrt{J^{(2)} \Om^{(2)}} \, ( k + \hlf ) \, , 
\label{DeigEnk}  \\
\ket{\chi_k^{D}} &\approx& \mathcal{N} 
\sum_j (2^k k!)^{-1/2} e^{- \xi_j^2/2 } H_k(\xi_j) \ket{1_j^D} \, , 
\label{DeigStk}
\eesa
where energies $E_k^D$ are relative to the dimer internal energy 
$(U - J^{(2)})$, $\xi_j = j \sqrt[4]{\Om^{(2)}/J^{(2)}} 
= j \sqrt[4]{\Om |U|/ J^2}$ is the discrete coordinate,  
and $\ket{1_j^D} \equiv \ck_j \ket{0}$ denotes a state with 
a single dimer at site $j$; obviously $\ket{1_j^D} = \ket{2_j}$.
The modified Bloch band $-2J^{(2)} \leq E_k^D \leq 2J^{(2)}$ for the dimer 
is restricted to the sites with
\be 
|j| \leq j_{\textrm{max}}^D \equiv \sqrt{\left( 1 + \frac{1}{\sqrt{2}} \right) 
\frac{J^{(2)}}{\Om^{(2)}} } \simeq 1.3 \sqrt{\frac{J}{\Om} \, \frac{J}{|U|}} \, ,
\ee 
thus containing $N^D = 2\lfloor j_{\textrm{max}}^D \rfloor +1$ energy 
levels $E_k^D$ with $0 \leq k < N^D$. The effective harmonic oscillator
frequency at the bottom of the modified Bloch band is 
$\hbar \om^D = 2 \sqrt{J^{(2)} \Om^{(2)}}$ and the dimer effective mass 
$\mu^D = \hbar^2/(2 J^{(2)}  d^2)$ is large ($J^{(2)} \ll J$) and positive. 
We have verified these conclusions by numerically solving the 
Schr\"odinger equation using the exact Hamiltonian (\ref{BHHam}) 
with the initial conditions corresponding to eigenstates (\ref{DeigStk})
of the effective Hamiltonian (\ref{DimHam}). As an example, in 
Fig. \ref{fig:adimdyn} we show the time evolution, or nearly complete 
absence thereof, of the system in the ground state of (\ref{DimHam}),
\be
\ket{\chi_0^{D}} \simeq \sqrt[8]{\frac{\Om^{(2)}}{\pi^2 J^{(2)}}}
\sum_j e^{- \xi_j^2/2 } \ket{1_j^D} 
= \sqrt[8]{\frac{\Om |U|}{\pi^2 J^2}} \sum_j e^{- \xi_j^2/2 } \ket{2_j} \, ,  
\label{aDgrSt}
\ee
with energy $E_0^D = -2J^{(2)} + \sqrt{J^{(2)} \Om^{(2)}}$. 

\begin{figure}[t]
\includegraphics[width=0.42\textwidth]{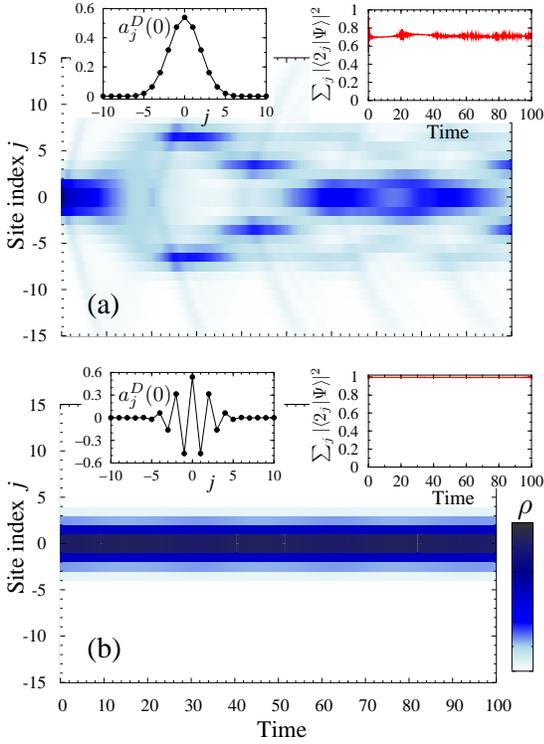}
\caption{Time evolution of density $\rho_j \equiv \expv{\hn_j}$ 
for repulsively-bound dimer in a combined periodic and parabolic 
potential with $J/\Om = 140$ and $U = 10 J$.
(a) Initial state $\ket{\Psi(0)}$ is the ground state $\ket{\chi_0^{D}}$ 
of attractive dimer, Eq.~(\ref{aDgrSt}), with amplitudes $a_j^D(0)$ 
shown in the left inset. 
(b) Initial state $\ket{\Psi(0)}$ is the ground state 
$\ket{\tilde{\chi}_0^{D}}$ of repulsive dimer, Eq.~(\ref{rDgrSt}), 
with amplitudes $a_j^D(0)$ shown in the left inset. 
Right insets are the projections $\sum_j |\braket{2_j}{\Psi}|^2$.}
\label{fig:rdimdyn}
\end{figure}

We next turn to the case of strong repulsive interaction $U > 0$.
The dimer tunneling constant is negative, $J^{(2)} < 0$, corresponding
to a negative effective mass $\mu^D$ \cite{PSAF}. As a result, 
$\ket{\chi_0^{D}}$ in Eq.~(\ref{aDgrSt}) is no longer the ground 
state of Hamiltonian (\ref{DimHam}), as attested in Fig.~\ref{fig:rdimdyn}(a).
Rather, it is a highly excited state. To see this, consider for a moment
a single particle in a flat lattice of $\bar{N}$ sites with $J < 0$. 
It follows from Eqs.~(\ref{EfinBB}), (\ref{psifinBB}) that the lowest 
energy state with $\bar{E}_{\bar{N}-1} = - 2 J \cos[\pi \bar{N}/(\bar{N}+1)] 
= - 2 |J| {\cos[\pi /(\bar{N}+1)]}$ is   
\be
\ket{\bar{\chi}_{\bar{N}-1}} = - \mathcal{N}  
\sum_{l=1}^{\bar{N}} \sin \left[ \frac{l \pi }{\bar{N}+1}\right]
e^{i l \pi} \ket{1_l} \, .
\ee  
Thus, in the limit of infinite lattice $\bar{N} \to \infty$, the ground 
state corresponds to the Bloch wave with quasi-momentum $q = \pi$.
Returning back to the repulsively-bound dimer in the combined periodic
and parabolic potential, we find that the low-energy eigenvalues 
are those of Eq.~(\ref{DeigEnk}) with the replacement $J^{(2)} \to |J^{(2)}|$,
while the corresponding eigenstates are given by 
\be
\ket{\tilde{\chi}_k^{D}} \approx \mathcal{N} 
\sum_j (2^k k!)^{-1/2} e^{- \xi_j^2/2 } H_k(\xi_j)  e^{i \pi j} \ket{1_j^D} \, .
\label{rDeigStk}
\ee
The ground state with $E_0^D = -2|J^{(2)}| + \sqrt{|J^{(2)}| \Om^{(2)}}$ is then
\bea
\ket{\tilde{\chi}_0^{D}} &\simeq & \sqrt[8]{\frac{\Om^{(2)}}{\pi^2 |J^{(2)}|}}
\sum_j e^{- \xi_j^2/2 } e^{i \pi j} \ket{1_j^D} 
\nonumber \\ 
&=& \sqrt[8]{\frac{\Om |U|}{\pi^2 J^2}} \sum_j e^{- \xi_j^2/2 } (-1)^j \ket{2_j} 
\, ,  \label{rDgrSt}
\eea
which is confirmed by our numerical simulations illustrated in 
Fig.~\ref{fig:rdimdyn}(b). Remarkably, the repulsive dimer 
appears to be tighter bound than the attractive one. 
The symmetry between the cases of $U < 0$ and $U > 0$ is broken 
due to the presence of a parabolic potential. 

\begin{figure}[t]
\includegraphics[width=0.42\textwidth]{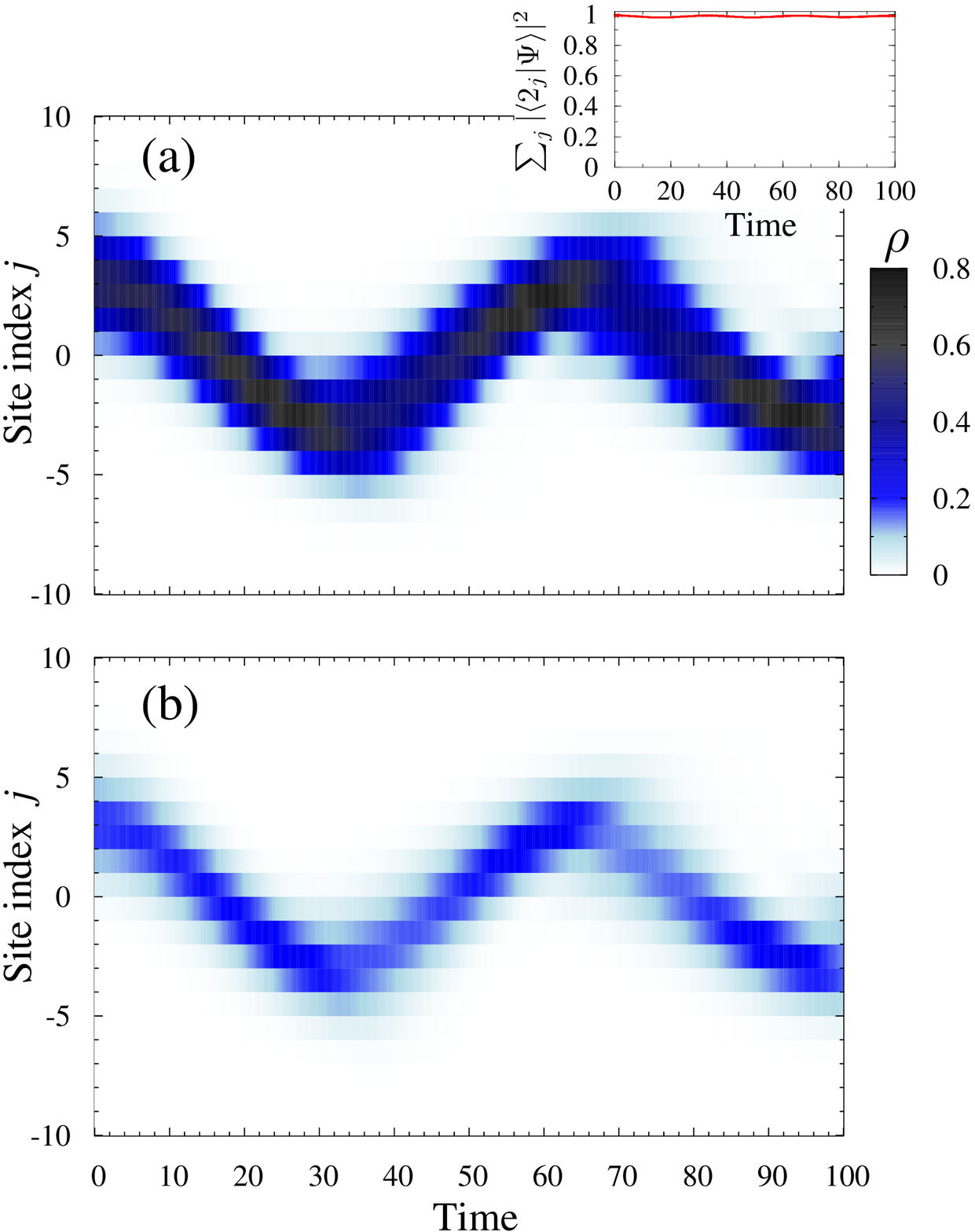}
\caption{Time evolution of (a) atom density $\rho_j \equiv \expv{\hn_j}$,
and (b) dimer density $\rho_j^D \equiv \expv{\hmd_j} \simeq \rho_j/2$ in a 
combined periodic and parabolic potential with $J/\Om = 140$ and $U = 10 J$.
Initial state $\ket{\Psi(0)}$ corresponds to the dimer ground state
$\ket{\tilde{\chi}_0^{D}}$ shifted by 3 sites from the trap center.
(a) is the numerical solution of the Schr\"odinger equation 
with the exact Hamiltonian~(\ref{BHHam}), while (b) is obtained 
with the effective Hamiltonian~(\ref{DimHam}). 
Inset in (a) is the projections $\sum_j |\braket{2_j}{\Psi}|^2$.}
\label{fig:dimdyn}
\end{figure}

Finally, in Fig.~\ref{fig:dimdyn} we show the dynamics of a dimer 
wavepacket $\ket{\Psi}$, represented by the ground state  
$\ket{\tilde{\chi}_0^{D}}$ initially shifted by $3$ sites from 
the trap center (for these parameters, $j_{\textrm{max}}^D \simeq 4.9$). 
Our simulations using the exact Hamiltonian (\ref{BHHam}) 
and the effective Hamiltonian (\ref{DimHam}) yield practically identical
results, which amount to periodic oscillations of the dimer wavepacket 
between the two sides of parabolic potential with period 
$\tau^D \simeq 2 \pi /\om^D = (\pi \hbar/ 2 J)  \sqrt{U/ \Om}$. 
Numerical simulations for attractively bound dimers reveal similar behaviour
but with considerably larger admixture of the single-particle 
states, $\sum_j |\braket{1_j}{\Psi}|^2 \lesssim 0.2$. This is 
another manifestation of the fact that the repulsive dimer in a 
combined periodic and weak parabolic potential is bound tighter  
than the attractive dimer under the otherwise similar conditions. 

\section{Conclusions}

To summarize, in this paper we have studied coherent quantum dynamics of 
one and two bosonic particles in a combined tight-binding periodic and 
shallow parabolic potential. Our studies are relevant to current experiments
with cold alkali atoms in optical lattices and weak magnetic (or optical) 
traps \cite{OptLatRev}. After revisiting the single-particle problem, 
we considered effective interaction-bound dimers recently realized in 
the experiment \cite{KWEtALPZ} with strong repulsive atom-atom interactions. 
We examined both cases of repulsively-bound and attractively-bound dimers
and identified similarities as well as marked differences in their static
and dynamic properties. In particular, a rather counterintuitive feature 
of the system revealed by the present work was that the repulsive dimers 
are bound stronger than the attractive dimers, as far as their 
ground states and coherent dynamics associated with low-energy states 
is concerned. In addition, we have shown that non-dispersive transport 
of carefully prepared atomic wavepackets can be achieved. As an extension
of the present work, we plan to study dimer--monomer resonant collisions 
and entanglement of the resulting wavepackets. Our results may be 
pertinent to quantum information schemes with cold atoms in optical 
lattices.

\acknowledgments

This work was supported by the 
EC Marie-Curie Research Training Network EMALI.

\end{document}